# Spin-controlled generation of indistinguishable and distinguishable photons from silicon vacancy centres in silicon carbide

## Authors

Naoya Morioka[1,2,*], Charles Babin[1], Roland Nagy[1], Izel Gediz[1], Erik Hesselmeier[1], Di Liu[1], Matthew Joliffe[1], Matthias Niethammer[1], Durga Dasari[1], Vadim Vorobyov[1], Roman Kolesov[1], Rainer Stöhr[1], Jawad Ul-Hassan[3], Nguyen Tien Son[3], Takeshi Ohshima[4], Péter Udvarhelyi[5,6,7], Gergő Thiering[6], Adam Gali[6,7], Jörg Wrachtrup[1], and Florian Kaiser[1,#]

\* n.morioka@pi3.uni-stuttgart.de

\# f.kaiser@pi3.uni-stuttgart.de

## Affiliations

1. 3rd Institute of Physics, University of Stuttgart and Institute for Quantum Science and Technology IQST, 70569, Stuttgart, Germany
2. Advanced Research and Innovation Center, DENSO CORPORATION, 470-0111, Nisshin, Japan
3. Department of Physics, Chemistry and Biology, Linköping University, SE-58183, Linköping, Sweden
4. National Institutes for Quantum and Radiological Science and Technology, 370-1292, Takasaki, Japan
5. Department of Biological Physics, Eötvös University, Pázmány Péter sétány 1/A, H-1117 Budapest, Hungary
6. Wigner Research Centre for Physics, P.O. Box 49, H-1525 Budapest, Hungary
7. Department of Atomic Physics, Budapest University of Technology and Economics, Budafoki út 8., H-1111 Budapest, Hungary

# Abstract

Quantum systems combining indistinguishable photon generation and spin-based quantum information processing are essential for remote quantum applications and networking. However, identification of suitable systems in scalable platforms remains a challenge. Here, we investigate the silicon vacancy centre in silicon carbide and demonstrate controlled emission of indistinguishable and distinguishable photons via coherent spin manipulation. Using strong off-resonant excitation and collecting photons from the ultra-stable zero-phonon line optical transitions, we show a two-photon interference contrast close to 90% in Hong-Ou-Mandel type experiments. Further, we exploit the system's intimate spin-photon relation to spin-control the colour and indistinguishability of consecutively emitted photons. Our results provide a deep insight into the system's spin-phonon-photon physics and underline the potential of the industrially compatible silicon carbide platform for measurement-based entanglement distribution and photonic cluster state generation. Additional coupling to quantum registers based on recently demonstrated coupled individual nuclear spins would further allow for high-level network-relevant quantum information processing, such as error correction and entanglement purification.

# Main text

## Introduction

The rise of quantum networks depends crucially on scalable quantum systems that combine quantum memories with long coherence times[1,2] and stable optical emission to mediate entanglement via interference of indistinguishable photons[3–8]. In this regard, solid state quantum systems are very appealing thanks to well-established fabrication techniques that promise chip-based integration and mass production[9]. A caveat coming along with many solid state systems is strong coupling to charge fluctuations, which leads to large spectral diffusion and ionisation, especially under off-resonant laser excitation and in surface proximity[10,11]. Recently, spectral stability has been greatly improved for divacancy centres in charge-depleted semiconductor diode structures, albeit ionisation remains an issue[12]. Alternatively, one can resort to systems with inversion symmetry, *e.g.* with germanium-vacancy[13,14], tin-vacancy centres[15], and silicon-vacancy[16] centres in diamond. For the latter, practical multi-millisecond spin coherence times have been demonstrated at ultracold temperatures in the millikelvin range[17].

Here, we use semiconductor colour centres that provide a naturally stable spin-photonic interface at temperatures up to 6.6 K and demonstrate spin-controlled quantum optical interference of indistinguishable and distinguishable photons. This milestone experiment demonstrates that the system possesses the necessary prerequisites for remote entanglement generation as targeted in scalable quantum networks.

Our technological platform is silicon carbide (SiC), which has recently acquired great interest in the quantum community as it is industrially compatible and hosts several appealing optically addressable spin-active quantum defects[18–20]. As shown in Fig. 1(a), we investigate here the negatively charged silicon vacancy centre at hexagonal lattice site (h-$V_{Si}$) in the common 4H polytype of SiC[21]. Despite the lack of inversion symmetry, it was recently shown that the optical transition of h-$V_{Si}$ is decoupled from charge fluctuations through identical wavefunction symmetries in the ground and V1 excited states[22]. Thanks to weak phonon coupling in the ground state, millisecond spin coherence times are achieved[20].

## Results

Our 4H-SiC host crystal is an isotopically purified (0001) epitaxial layer ($^{28}$Si ~ 99.85%, $^{12}$C ~ 99.98%.), which is irradiated with 2 MeV electrons to generate isolated single h-$V_{Si}$ centres. The crystal is slightly n-type to ensure that h-$V_{Si}$ centres are in the desired negatively charged state (see Methods). As shown in Fig. 1(b), the h-$V_{Si}$ centre presents a spin quartet system ($S = \frac{3}{2}$). At zero external magnetic field, spin sublevels $m_S = \pm\frac{1}{2}$ and $m_S = \pm\frac{3}{2}$ are pairwise degenerate in the ground state (GS) and the V1 excited state (ES). As the GS zero field splitting is relatively small at 4.5 MHz[21], we apply a magnetic field ($B \approx 9$ G) along the system's symmetry axis (c-axis). This lifts the degeneracy and suppresses parasitic spin mixing due to external stray fields. Optical transitions are linearly polarised and associated with two zero phonon line (ZPL) transitions at 861.7 nm. They are assigned to $|\pm 1/2\rangle_{GS} \leftrightarrow |\pm 1/2\rangle_{ES}$ and $|\pm 3/2\rangle_{GS} \leftrightarrow |\pm 3/2\rangle_{ES}$ transitions, and labelled $A_1$ and $A_2$, respectively. The energy separation between the two transitions is about 1 GHz, which is mainly determined by the ES zero field splitting[21]. It was recently shown that all optical transitions are fully spin conserving, which provides several pathways for spin-photon entanglement generation[9,23]. However, spin-flip processes can still be mediated by non-radiative intersystem crossing involving metastable states, which allows *e.g.* for deterministic spin state initialisation[21,24]. Note that we do not consider the system's second excited state, V1', whose fluorescence at 858 nm is strongly reduced at cryogenic temperatures due to ultrafast relaxation to the V1 state[20]. Thus, and as shown in Fig. 1(c), off-resonant excitation results in photoluminescence in the V1 ZPL or the associated phonon side band (PSB) with 6 ns lifetime[20].

Single h-$V_{Si}$ centres are addressed via confocal microscopy (see Methods). Fig. 1(d) shows the fine structure of the V1 ZPL emission, recorded at a temperature of $T = 5.0$ K. To this end, we use continuous-wave off-resonant excitation at 730 nm with an optical power of about 2 mW, which is almost 10 times higher than the saturation power[20]. As a high-resolution spectrometer, we use a home-built tunable Fabry-Pérot filter cavity with a linewidth of $40 \pm 2$ MHz. By integrating the emission spectrum over 20 minutes, we clearly resolve the two optical lines corresponding to the spin-conserving $A_1$ and $A_2$ transitions. Surprisingly, even under strong off-resonant excitation, the deconvoluted linewidths remain very close to the lifetime limit (~27 MHz), *i.e.* $89 \pm 20$ MHz and $51 \pm 14$ MHz for the $A_1$ and $A_2$ transition, respectively.

Aside from superior spectral purity and robustness, perfect two-photon interference crucially requires the interfering photons to be in pure single-photon states, *i.e.* not degraded by spurious multi-photon contributions[25]. To evaluate the quality of our quantum light source, we use pulsed off-resonant excitation at 780 nm (PicoQuant LDH-P-C-780, pulse length: 56 ps, repetition rate: 20.5 MHz), and record the second-order autocorrelation function $g^2(\tau)$ of ZPL photons in a standard Hanbury Brown and Twiss arrangement[26]. As shown in Fig. 1(e), we observe a strong antibunching at zero time delay, clearly evidencing that the defect is a single photon source ($g^2(\tau = 0) < 0.5$). Fig. 1(f) shows the recorded value of $g^2(\tau = 0)$ as a function of the laser pulse energy above saturation (4 pJ, see Supplementary Note 1). The increase of $g^2(\tau = 0)$ is mainly due to noise from background fluorescence below 6 pJ (see Supplementary Note 3). Above 6 pJ, the increased probability to induce two optical excitations during one laser pulse degrades $g^2(\tau = 0)$. However, under all conditions, we observe $g^2(\tau = 0) < 0.25$ without any background subtraction, which underlines the h-$V_{Si}$ centre's single-photon emission quality.

To prove that the h-$V_{Si}$ centre generates streams of indistinguishable photons in the $A_1$ and $A_2$ transitions, we perform a Hong-Ou-Mandel (HOM) interference[27] experiment on two consecutively emitted photons[28]. Fig. 2(a) shows the related setup schematically. We use the off-resonant picosecond laser to excite the single defect twice with an interval of $\delta t = 48.7$ ns, and we repeat this sequence every $10 \cdot \delta t = 487$ ns. The laser pulse energy is set to 5.5 pJ, corresponding to about 74% excitation probability per pulse (see Supplementary Note 1). The ZPL emission is coupled into a single-mode-fibre based unbalanced Mach-Zehnder interferometer with a path travel time difference of $\delta t = 48.7$ ns. The interferometer's outputs are connected to superconducting nanowire single-photon detectors (SNSPD, Photon Spot Inc.) with sub-ns time resolution. We record the time differences between both SNSPD detection events using a time tagger (Swabian Instruments). The timing jitter of our detection system is about 0.4 ns.

Let us assume now that two consecutive ZPL photons enter the interferometer during one experimental sequence (*i.e.* no non-radiative intersystem crossing occurred). The most interesting case occurs when the early and the late photons take the long and short interferometer paths, respectively, and thus arrive simultaneously at the output beam-splitter (BS$_2$) through different inputs. Provided that both photons are indistinguishable, HOM interference occurs and they must leave BS$_2$ as a pair through the same output port[27]. Experimentally, this is measured through the reduction of coincidence events at zero time delay between both SNSPDs. Note that the normalised coincidence rate reduction is a direct measure of the photon indistinguishability. The normalisation process is actually implemented automatically in our setup. Namely, 25% of the successfully created photon pairs take "wrong" opposite paths, *i.e.* the early and the late photons take the short and long interferometer paths, respectively. In this case, they arrive at BS$_2$ with a time difference of $\pm 2 \cdot \delta t$. As this delay exceeds the excited state lifetime (6 ns), no interference is observed. Similarly, the remaining 50% of the paired photons choose identical interferometer paths and arrive at BS$_2$ without interference with a time difference of $\pm \delta t$. Thus, we expect that the five detection rates associated with the different coincidence time differences $A_{-2 \cdot \delta t}$, $A_{-\delta t}$, $A_0$, $A_{+\cdot \delta t}$, and $A_{+2 \cdot \delta t}$ show the well-known ratio of 1:2:0:2:1.[28]

Fig. 2(b) shows typical raw data for the HOM experiment. The central peak at $\tau = 0$ ns is strongly suppressed compared to the side peaks at $\tau = \pm \delta t$ and $\tau = \pm 2 \cdot \delta t$. Each peak exhibits an exponential decay of 6 ns corresponding to the V1 excited state lifetime. Following the analysis by Santori et al.[28], we determine the raw HOM interference visibility $V_0 = 1 - 2 \cdot \frac{A_0}{A_{-\delta t} + A_{+\delta t}} = 0.69 \pm 0.02$. Since this value greatly exceeds $2 \cdot g^{(2)}(\tau = 0)$, generation of two-photon entanglement can be straightforwardly implemented[29–31].

As mentioned earlier (see Figs. 1(b) and (d)), optical transitions are spin conserving and intimately linked to spin sublevels. *I.e.* $|\pm 1/2\rangle_{GS} \leftrightarrow |\pm 1/2\rangle_{ES}$ levels are connected via the $A_1$ optical transition, and $|\pm 3/2\rangle_{GS} \leftrightarrow |\pm 3/2\rangle_{ES}$ via $A_2$. Consequently, without intersystem crossing, repeated off-resonant excitation leads to streams of consecutively emitted indistinguishable "red" photons in the $A_1$ or "blue" photons in the $A_2$ transition, respectively. As one of the system's key assets, we highlight that the h-$V_{Si}$ centre in SiC possesses a highly coherent ground state spin quartet[20,21]. This provides a direct control mechanism of the colour of consecutively emitted photons. *E.g.* by flipping the ground state spin from the subspace $|\pm 1/2\rangle_{GS}$ to $|\pm 3/2\rangle_{GS}$, the first and second photon will be "red" and "blue", respectively. Fig. 3(a) shows the experimental sequence. The four ground state spin levels are non-degenerate due to the external magnetic field. After the first excitation, we apply a short radiofrequency wave (RF) pulse at a frequency of 30.3 MHz. Although the pulse frequency is centred at the transition between $|+1/2\rangle_{GS}$ and $|+3/2\rangle_{GS}$, it manipulates the spin state between $|\pm 1/2\rangle_{GS}$ and $|\pm 3/2\rangle_{GS}$ subspaces with about 70% fidelity due to the high power and the short duration (for more details, see Supplementary Note 4). Fig. 3(b) shows the resulting HOM interference pattern at RF power of 30 dBm and a pulse length of 19 ns, which corresponds to a $\pi/2$-pulse (see Supplementary Note 4). The coincidence peak at zero time delay reappears clearly, due to the emission of distinguishable photons. Additionally, as shown in Fig. 3(c), the interference pattern around zero time delay shows the expected modulation with a frequency of $\approx 1$ GHz, matching the frequency difference between $A_1$ and $A_2$ optical transitions[32]. We note that the high RF pulse power causes significant optical linewidth broadening due to thermal heating, limiting the maximum HOM visibility to $0.56 \pm 0.04$ and $V_{max} = 0.65 \pm 0.05$ after correcting for experimental imperfections (as discussed later and in Supplementary Note 4). By normalising the recorded and corrected visibility $V_{measured}$ to $V_{max}$, we obtain a corrected HOM visibility of $V_{norm, \pi/2} = \frac{V_{measured}}{V_{max}} = 0.61 \pm 0.10$, which matches the expectation for a $\pi/2$-pulse (for details, see Supplementary Note 4). To prove that our spin-to-photon interface is indeed based on coherent spin manipulation, we repeat the experiment with two additional RF pulse lengths of 10 ns and 29 ns (corresponding to a $\pi/4$-pulse and $3\pi/4$-pulse, respectively). The observed visibilities are $V_{norm, \pi/4} = 0.89 \pm 0.10$ and $V_{norm, 3\pi/4} = 0.37 \pm 0.26$, respectively. Data for the three measurements above is shown in Fig. 3(d). The data matches the theoretical model very precisely, which is based on independently measured Rabi oscillations (see Supplementary Note 4). Consequently, the results corroborate that the photonic emission of the h-$V_{Si}$ centre can be controlled by coherent spin manipulation, which is crucial for remote entanglement generation[5–8,33,34].

In the last step, we provide a deeper insight into the system's temperature-dependent spin-phonon-photon physics. For this, we consider the visibility reduction in the raw data of the HOM experiment without RF pulses. To explain deviations from non-unity visibility, we consider six detrimental factors: (i) pure dephasing in the excited state due to phonon scattering[35], (ii) spectral diffusion due to local charge redistribution[36], (iii) two-photon emission during one laser pulse[28,37], (iv) background counts, *e.g.* laser breakthrough, bulk and surface fluorescence, and Raman scattering, (v) interferometer imperfections, *e.g.* non-unity fringe contrast and unbalanced beam splitter transmissivity and reflectivity[28], and (vi) the photon arrival timing jitter at BS$_2$, *e.g.* due to finite laser pulse duration[38] as well as timing jitter of the laser. We infer that factors (v) and (vi) amount to 1% and 0.8% contrast reduction, respectively (see Supplementary Note 2 and 3). Experimentally, we minimise the fast components of factors (iii) and (iv) by time-gating the coincidence detection, *i.e.* detector clicks are only considered posterior to a delay time $t_{\text{Start}}$ (see Methods). Generally, we find that the visibility saturates for $t_{\text{Start}} > 1.5$ ns, indicating that laser-induced noise is efficiently filtered out at this point. To infer the contribution of pure dephasing and spectral diffusion, we extend the analysis by A. Thoma et al.[35] and include coincidence window gating with start and stop delays $t_{\text{Start}}$ and $t_{\text{Stop}}$, respectively (for more details, see Supplementary Note 5). This provides us with an additional tuning knob, as we can now actively control the time window $\Delta t = t_{\text{Stop}} - t_{\text{Start}}$ in which the HOM interference contrast is sensitive to pure dephasing (see Fig. 4(a)). As an example, Fig. 4(b) shows the HOM pattern at $T = 5.0$ K for $t_{\text{Start}} = 3.5$ ns and a short time window $\Delta t = 4$ ns. For those settings, we obtain an uncorrected raw visibility of $V_{0,\text{gated}} = 0.85 \pm 0.04$. For arbitrary time gating settings, we find an analytic expression for the expected HOM visibility:

$$V = \frac{1}{(1 - e^{-\Gamma \cdot \Delta t})^2} \left[ \frac{\Gamma}{\Gamma + \gamma} + \frac{\Gamma}{\Gamma - \gamma} e^{-2\Gamma \cdot \Delta t} - \frac{2\Gamma^2}{\Gamma^2 - \gamma^2} e^{-(\Gamma + \gamma) \Delta t} \right]. \qquad (1)$$

Here, $\Gamma = \frac{1}{6 \text{ ns}}$ is the inverse excited state lifetime, and $\gamma = \Gamma'_0 \left[ 1 - e^{-(\delta t / \tau_c)^2} \right] + 2\gamma'$, with $\Gamma'_0$ being the amplitude of spectral diffusion, $\tau_c$ the associated time constant, and $\gamma'$ the pure dephasing rate of the single emitter. To infer $\gamma$, the experimental data has to be corrected for experimental imperfections (interferometer, finite SN, and photon arrival timing jitter). Fig. 4(c) shows the corrected HOM visibility as a function of $\Delta t$ for fixed $t_{\text{Start}} = 3.5$ ns. The measurements are performed by adjusting the cryostat to three different temperatures $T = 5.0$ K, 5.9 K and 6.8 K. By fitting the data with the model in Eq. (1), we extract $\gamma$. We note that one contribution to spectral diffusion is thermal ionization of nearby charge traps, usually occurring at several tens of Kelvin[39–41]. Thus, in our experiments, spectral diffusion is attributed to laser ionization, which depends very weakly on temperature and occurs usually at a microsecond time scales[41]. Consequently, we have $\tau_c \gg \delta t$, such that HOM visibility reduction is essentially only limited by pure dephasing ($\gamma'$). This permits us to provide an upper limit for the pure dephasing rate $\gamma'_{\text{max}}$ (obtained from the "fast" HOM experiments), as well as the associated amplitude of spectral diffusion $\Gamma'_0$ (obtained from "slow" absorption linewidth measurements). The results of the temperature dependent measurements are summarised in Fig. 4(d). As expected, we find that $\Gamma'_0$ remains nearly constant over the measured temperature range. The increase of $\gamma'_{max}$ is described by vibronic interaction of the V1 and V1' excited states[42], which are separated by a relatively small energy gap of $\Delta E = 4.4$ meV. As we detail in the Supplementary Note 6, at the experimental temperatures, single acoustic phonon scattering processes cause pure dephasing with a rate given by

$$\gamma'_{max}(T) = A \cdot \frac{(\Delta E)^3}{e^{\Delta E/k_B \cdot T} - 1}. \tag{2}$$

Here, $k_B$ is Boltzmann's constant, and the prefactor $A$ describes the phonon interaction strength. By fitting the data, we obtain $A = 2\pi \cdot (365 \pm 36)$ MHz $\cdot$ (meV)$^{-3}$. Interestingly, we find that pure dephasing for the h-$V_{Si}$ centre is comparatively low at cryogenic temperatures. *E.g.* considering photonic interference experiments without any time-gating, pure dephasing reduces the optical coherence time to half of the transform limit ($2\Gamma^{-1} = 12$ ns) only at $T_{crit} > 6.6$ K. This temperature is comparable to quantum dot single-photon sources[43], and conveniently surpassed with standard cryostat equipment. Below $T_{crit}$, the remaining small emission linewidth broadening is mainly due to slow spectral diffusion. Although this does not present a significant influence for HOM experiments, it may be even further suppressed with improved crystal growth and associated annealing procedures to reduce charge traps. Alternatively, electronic device structures might be promising to control the charge environment[12,44–46].

## Conclusion

In summary, we have demonstrated spin-controlled generation of indistinguishable and distinguishable photons from a single h-$V_{Si}$ centre in SiC. Despite performing our experiments under strong off-resonant excitation, pure dephasing and spectral diffusion are exceptionally low, such that the quality of our raw data is sufficient for photonic entanglement generation[29–31]. Our semiconductor platform has thus demonstrated a level of spectral stability which is on par with inversion-symmetry defects in diamond[13–16], quantum dots[35,47], and single molecules[31]. Furthermore, we showed that the system possesses an intimate spin-photon interface through which we deterministically tuned the degree of photon indistinguishability via coherent spin manipulation. Note that the system's millisecond spin coherence times[21] are sufficient for high-level spin-photon manipulation, such as remote entanglement generation[5–7] in a scalable quantum repeater network[8]. Implementing the related protocols necessitates optimised RF driving to reduce heating issues, which will suppress pure dephasing to negligible levels. To provide the necessary optical phase reference, resonant excitation is required, which requires several orders of magnitude lower optical powers, such that spectral diffusion will also be greatly suppressed[48]. In the perspective of quantum repeaters, long storage time quantum memories can be conveniently realised with recently demonstrated coupling to individual nuclear spins[21]. Implementation of state-of-the-art SiC photonic nanostructures would further increase light collection efficiency and ultra-compact chip-based integration[49], such that it will become realistic to use our system for generation of high-photon-number cluster states[50].

## Acknowledgements

We warmly thank Christoph Becher, Jürgen Eschner and Mete Atatüre for fruitful discussions regarding data analysis. We thank Andy Steinmann, Mario Hentschel, Jonas Meinel, Stephan Hirschmann, Ilja Gerhardt, Timo Görlitz, Matthias Widmann, Torsten Rendler, Yu-Chen Chen, Thomas Kornher, Kai-Mei Fu, Karten Frenner, Photon Spot Inc., PicoQuant GmbH, and Swabian Instruments GmbH for technical help.

N.M., C.B., R.N., I.G., E.H., D.L., M.J., M.N., D.D., V.V., R.K., R.S., J.W. and F.K. acknowledge support by the European Research Council (ERC) grant SMel, the European Commission Marie Curie ETN "QuSCo" (GA No 765267), the Max Planck Society, the Humboldt Foundation, and the German Science Foundation (SPP 1601). R.N. acknowledges support by the Carl-Zeiss-Stiftung. N.T.S. acknowledges the Swedish Research Council (grant No. VR 2016-04068). J.U.H. thanks the Swedish Energy Agency (43611-1). N.T.S. and J.U.H. thank the Knut and Alice Wallenberg Foundation (Grant No. KAW 2018.0071). A.G., N.T.S. and J.U.H. thank the EU H2020 project QuanTELCO (Grant No. 862721). T.O. thanks the Japan Society for the Promotion of Science (Grant No. JSPS KAKENHI 17H01056 and 18H03770). A.G. acknowledges the National Excellence Program of Quantum-Coherent Materials Project (Hungarian NKFIH Grant No. KKP129866), the EU QuantERA Q-Magine Project (Grant No. 127889), the QuantERA Nanospin Project (Grant No. 127902), and the National Quantum Technology Program (Grant No. 2017-1.2.1-NKP-2017-00001). A.G. and J.W. acknowledge the EU-FET Flagship on Quantum Technologies through the project ASTERIQS. J.W. and F.K. acknowledge the EU-FET Flagship on Quantum Technologies through the project QIA.

## Author contributions

N.M., R.N., J.W. and F.K. conceived and designed the experiment; N.M., I.G., E.H., D.L., and F.K. performed the experiment; C.B., M.J., V.V., R.K., and R.S. provided experimental assistance; N.M., P.U., G.T., A.G., J.W., and F.K. analysed the data; J.U.H. and N.T.S. prepared and characterised SiC materials; T.O. performed electron beam irradiation; R.N. fabricated solid immersion lenses; N.M., I.G., and M.N. developed software for data acquisition and experimental control; N.M., C.B., D.L., D.D., V.V., P.U., G.T., A.G., J.W., and F.K. provided theoretical support. N.M., J.W., and F.K. wrote the manuscript. All authors provided helpful comments during the writing process.

# Figures

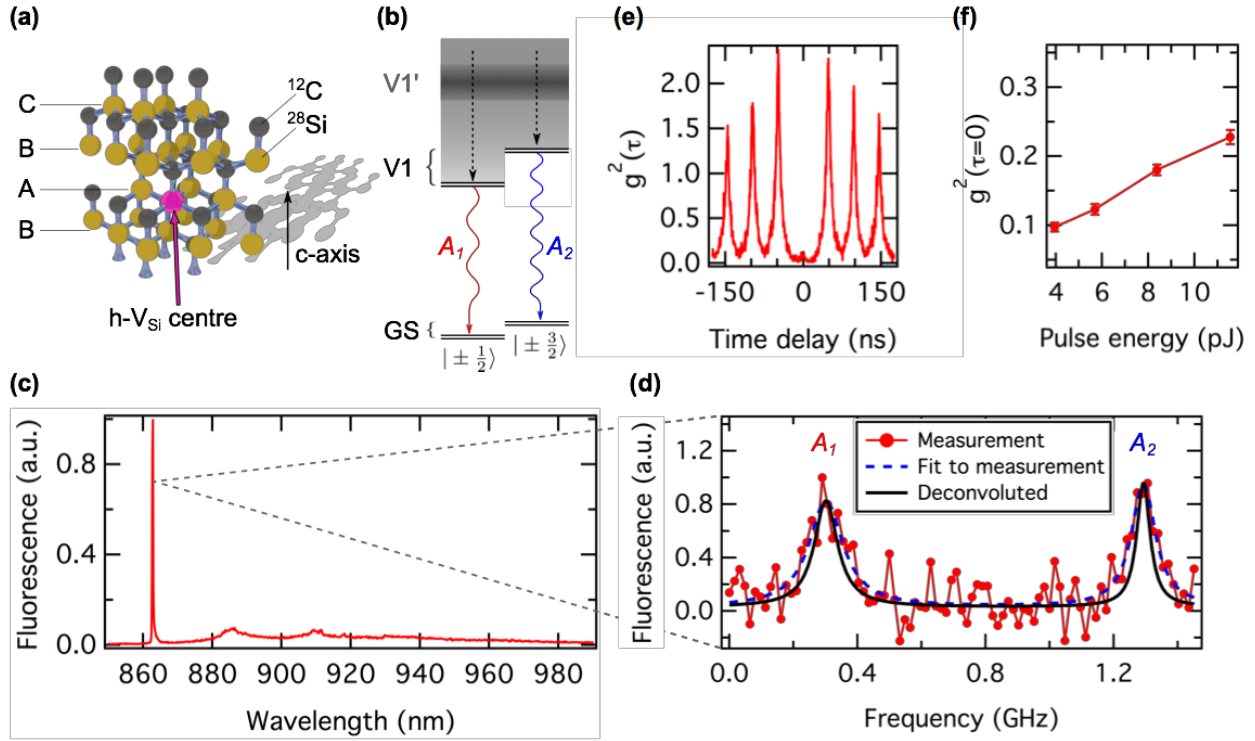

Figure 1. (a) Crystallographic structure of 4H-SiC and position of h-$V_{Si}$ centre (highlighted by a pink sphere symbolising a missing Si atom). (b) Level structure of the h-$V_{Si}$ centre at zero magnetic field. Ground state (GS) and V1 excited state show degenerate sublevels $m_S = \pm\frac{1}{2}$ and $m_S = \pm\frac{3}{2}$. Optical transitions between V1 and GS are spin conserving and associated with the transitions $A_1$ and $A_2$. Emission from the second excited state, V1', is not observed due to ultrafast relaxation (dashed arrows). (c) Single h-$V_{Si}$ centre emission spectrum under off-resonant excitation. (d) ZPL emission fine structure recorded over 20 minutes. The two emission lines associated with $A_1$ and $A_2$ transitions are clearly resolved. The blue dashed line is a Lorentzian fit to the raw data, giving linewidths (FWHM) of $129 \pm 20$ MHz and $91 \pm 14$ MHz for the $A_1$ and $A_2$ lines, respectively. After deconvolution correction for the finite linewidth of the scanning Fabry Pérot cavity (Lorentzian FWHM of $40 \pm 2$ MHz), the resulting real emission linewidths are $89 \pm 20$ MHz and $51 \pm 14$ MHz, respectively, which is very close to the Fourier transform limit. (e) Second-order autocorrelation function recorded for a single h-$V_{Si}$ centre under pulsed off-resonant excitation (pulse energy: 5.7 pJ). We observe $g^2(\tau = 0) = 0.12 \pm 0.01$, clearly indicating single-photon emission. (f) $g^2(\tau = 0)$ as a function of laser pulse energy. The line is a guide to the eye.

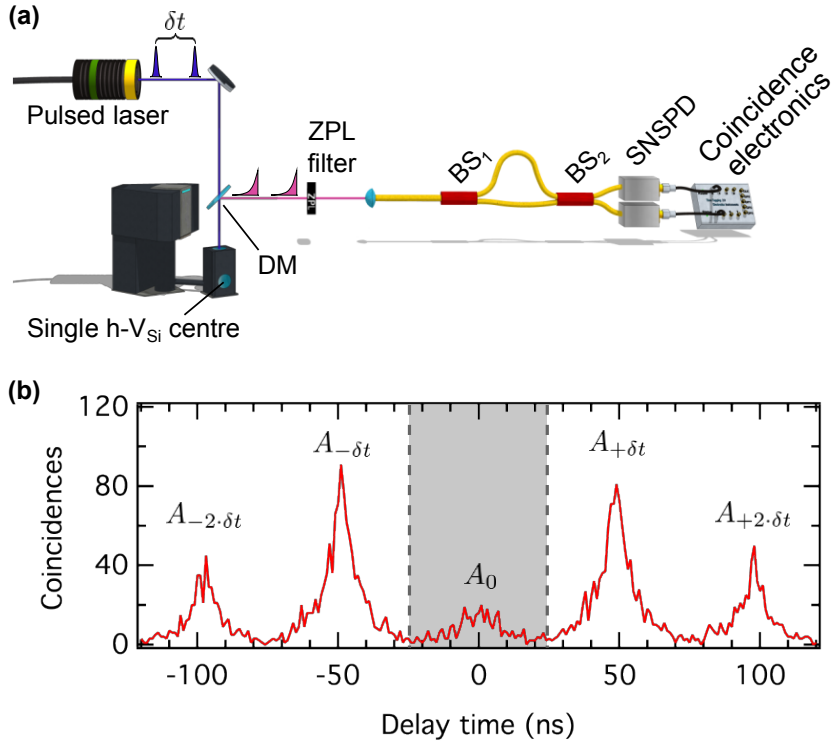

Figure 2. (a) Schematic setup for HOM interference with two photons from a single h-$V_{Si}$ centre in SiC. Two laser pulses excite the h-$V_{Si}$ centre with a time delay $\delta t = 48.7$ ns. Consecutively emitted ZPL photons are sent to an unbalanced Mach-Zehnder interferometer. SNSPD and coincidence electronics are used to record the two-photon statistics at the output. (b) Two-photon coincidence counts as a function of the detection time delay between both SNSPDs. The strongly suppressed peak at zero time delay witnesses Hong-Ou-Mandel interference. The grey area between dashed lines symbolises the integration time window that is used for evaluating the interference contrast $V_0 = 0.69 \pm 0.02$. The results shown are raw data without any correction.

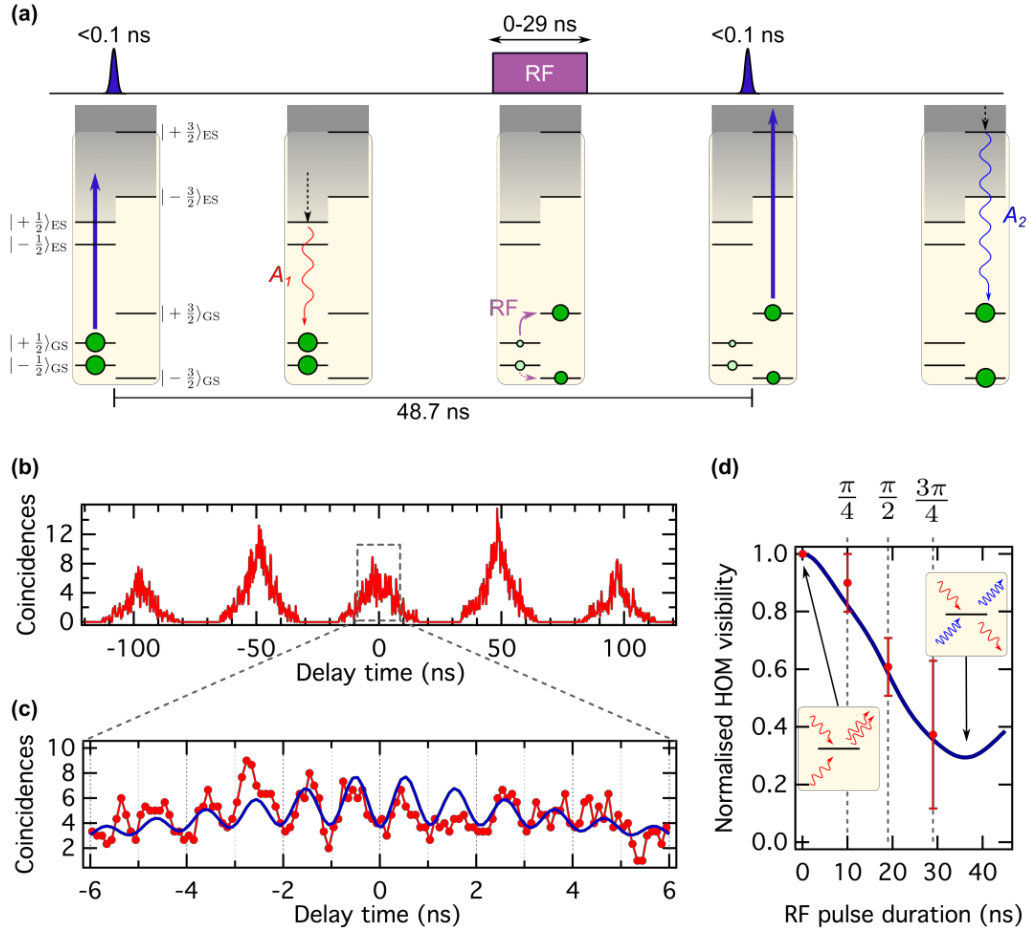

Figure 3. (a) Schematics for spin-controlled generation of distinguishable photons from a single h-$V_{Si}$ centre. The sketch shows the realisation when the system is initially in the $m_S = \pm\frac{1}{2}$ subspace. The first laser excitation results in a red photon ($A_1$ line). A subsequent RF pulse with variable duration transfers population (partially) from $m_S = \pm\frac{1}{2}$ to $m_S = \pm\frac{3}{2}$. The second excitation results in a blue photon ($A_2$ line), which makes the two interfering photons maximally distinguishable, and projects the system in to the $m_S = \pm\frac{3}{2}$ subspace. (b) Two-photon coincidence counts as a function of the delay time for a RF pulse duration of 19 ns (corresponding to a $\pi/2$-pulse). The coincidence peak at zero time delay reappears. Data are recorded at 0.1 ns timing resolution and averaged over three points to improve signal-to-noise. (c) Zoom-in of the HOM interference pattern revealing the fringe pattern with the expected modulation at $0.966 \pm 0.007$ GHz. Red dots are uncorrected data, and the blue line is the fit to the data (for details on the fit model, see Supplementary Note 7). (d) HOM contrast as a function of the RF pulse duration. Rabi-like oscillations are observed, demonstrating that coherent spin manipulation controls the degree of photon indistinguishability. Red dots are data and the blue line is the theoretical model considering independently measured Rabi oscillations (see Supplementary Note 4).

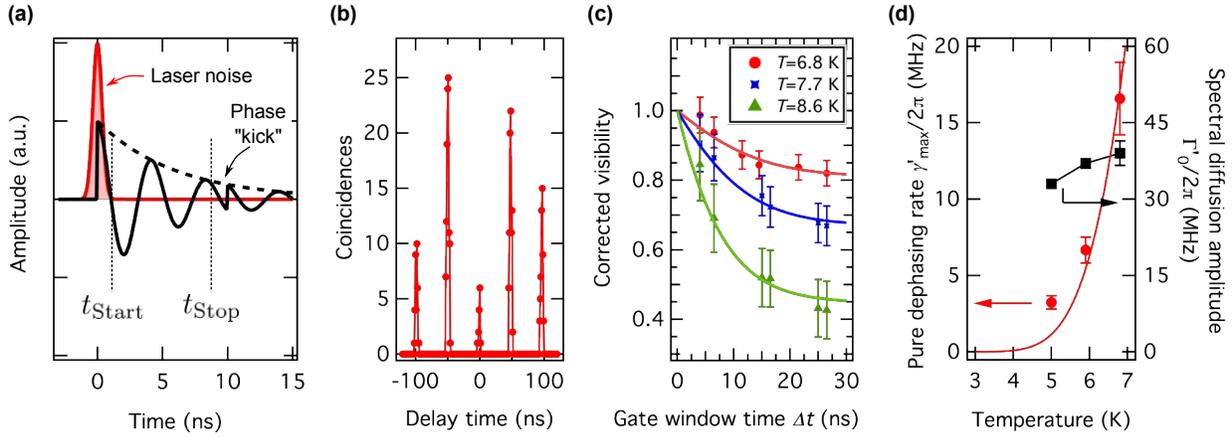

Figure 4. (a) Visualisation of the time-resolved HOM experiment (not to scale for better visualisation). The black solid line symbolises the single-photon wavefunction, while the black dashed line represents its envelope. For short times close to 0 ns, laser noise photons are observed (red pulse). This contribution is conveniently filtered out by only accepting detection events posterior to the start gate delay $t_{Start}$. In our visualisation, the photon wave function experiences a phase "kick" at a time delay of about 10 ns, due to phonon scattering in the V1 excited state. By having a variable stop gate time $t_{Stop}$, we effectively control the integration window $\Delta t = t_{Stop} - t_{Start}$. This allows us to infer the time scale of pure dephasing. (b) Uncorrected HOM pattern at $T = 5.0$ K for $t_{Start} = 3.5$ ns and $\Delta t = 4$ ns. Due to the short time gating, the raw visibility increases to $V_0 = 0.85 \pm 0.04$, which matches the theoretical expectation ($0.86 \pm 0.01$) considering only experimental imperfections (see Supplementary Note 3). Dots are data and lines are visual guides. (c) Corrected HOM visibility at $t_{Start} = 3.5$ ns and varying $\Delta t$ and at three different temperatures. For $\Delta t \to 0$ ns, visibilities approach unity, as pure dephasing is completely suppressed. Dots are data, and lines are fits according to Eq. (1). (d) Temperature-dependent pure dephasing rate $\gamma'_{max}$ and associated spectral diffusion amplitude $\Gamma'_0$. The red line is the fit considering the model described with Eq. (2). The black line is a guide to the eye.

## Methods

### Sample preparation

The 100 μm thick 4H-$^{28}$Si$^{12}$C silicon carbide layer is grown by chemical vapour deposition (CVD) on an n-type (0001) 4H-SiC substrate. The isotope purity is measured by secondary ion mass spectroscopy (SIMS) and inferred to be $^{28}$Si ~ 99.85% and $^{12}$C ~99.98%. Current-voltage measurements at room temperature shows that the layer is n-type with a free carrier concentration of ~ $6 \cdot 10^{13}$ cm$^{-3}$, which is close to the concentration of shallow nitrogen donors of ~ $3.5 \cdot 10^{13}$ cm$^{-3}$. Deep level transient spectroscopy measurements show that the dominant electron trap in the layer is related to the carbon vacancy with a concentration in the mid $10^{12}$ cm$^{-3}$ range. Minority carrier lifetime mapping of the carrier shows a homogeneous carrier lifetime of ~ 0.6 μs. We expect the real value to be twice as high, as an optical method with high injection was used[51]. Thus, the density of all electron traps should be limited to the mid $10^{13}$ cm$^{-3}$ range[52]. Individually addressable silicon vacancy centres were created through room temperature electron beam irradiation at 2 MeV with a fluence of $10^{12}$ cm$^{-2}$. Parasitic defects, such as carbon vacancies, interstitials, anti-sites and their associated defects, are expected to be below mid $10^{12}$ cm$^{-3}$. Some interstitial-related defects were removed by subsequent annealing at 300°C for 30 minutes.

To improve light extraction efficiency out of the high refractive index material ($n \approx 2.6$), we fabricate a solid immersion lens using a focused ion beam milling machine (Helios NanoLab 650). The related surface contaminations and modifications are subsequently removed by wet and dry etching treatments[19].

**Experimental setup**

All the experiments were performed at cryogenic temperatures in a Montana Instruments Cryostation. A home-built confocal microscope[20] was used for optical excitation and subsequent fluorescence detection of single silicon vacancies.

Continuous-wave off-resonant optical excitation of single silicon vacancy centres was performed with a 730 nm laser diode. Pulsed off-resonant excitation at 780 nm was performed using a picosecond laser diode (PicoQuant LDH-P-C-780). For resonant optical excitation at 861.7 nm we used an external cavity tunable diode laser (Toptica DLC DL PRO 850). All lasers are spatial-mode cleaned by coupling to a single-mode fibre. For the picosecond laser, we employ a 780 nm bandpass filter after the fibre output to suppress Raman noise. Light is focussed onto the sample with a vacuum-compatible microscope objective (Zeiss EC Epiplan-Neofluar 100×, NA=0.9).

Note that the used 4H-SiC sample was flipped to the side, *i.e.* by 90° compared to the c-axis, such that the polarisation of the excitation lasers is parallel to the c-axis ($E \| c$), which allows to excite the V1 excited state with maximum efficiency[20,53].

The fluorescence emission is collected by the same microscope objective and split by a dichroic mirror (Semrock Versa Chrome Edge) in two parts, *i.e.* zero-phonon line (ZPL at 861.7 nm) and phonon side band (PSB at 875 – 950 nm). PSB fluorescence is detected using a silicon avalanche photodiode (Excelitas SPCM-AQRH-W4).

ZPL emission is directed to a fully-fibred Mach-Zehnder type interferometer with a path length difference of $\approx 10$ m. A fibre polarisation controller in the long interferometer arm allows to match the photon polarisation rotation experienced in both arms. Both interferometer outputs are directed to superconducting nanowire single-photon detectors (SNSPD, PhotonSpot Inc.) with 80-85% detection efficiency and sub-Hz dark count rates.

In order to coherently manipulate ground state spin populations, microwaves are applied through a 50 μm thick copper wire spanned over the 4H-SiC sample in close proximity to the investigated h-$V_{Si}$ defect centre.

## Filter cavity for ZPL fine structure investigation

To investigate the fine structure of the ZPL, we use a home-built fibre-coupled plano-concave Fabry-Pérot cavity. The cavity has the following specifications: Free spectral range: 8.05 GHz; Finesse: $200 \pm 8$; Linewidth: $40 \pm 2$ MHz; Input-fibre-to-output-fibre transmission: 75%. The cavity length is tunable via a piezoelectric actuator on which one of the two mirrors is glued. The cavity housing is made from Invar to ensure good thermal stability. The typical drift of the cavity was measured to be on the order of 5 kHz/s.

## Time-gating scheme

To filter out noise components induced by the off-resonant high-power pulsed laser excitation and to perform time-resolved HOM experiments for investigation of pure dephasing, a time-gated photon detection scheme is implemented by software. All photon click signals from the SNSPDs are time tagged by coincidence electronics (Swabian Instruments Time Tagger 20) and referenced to the laser pulse timing. Via software postprocessing we then implement time-gating by validating only photon clicks between $t_{Start} \leq t \leq t_{Stop}$ (here, $t = 0$ represents the earliest possible photon arrival time). Due to the software implementation on the Time Tagger 20, the experimental data points with different settings of $(t_{Start}, t_{Stop})$ are obtained within a single measurement.

# Supplementary Information: Spin-controlled generation of indistinguishable and distinguishable photons from silicon vacancy centres in silicon carbide

## Authors

Naoya Morioka[1,2,*], Charles Babin[1], Roland Nagy[1], Izel Gediz[1], Erik Hesselmeier[1], Di Liu[1], Matthew Joliffe[1], Matthias Niethammer[1], Durga Dasari[1], Vadim Vorobyov[1], Roman Kolesov[1], Rainer Stöhr[1], Jawad Ul-Hassan[3], Nguyen Tien Son[3], Takeshi Ohshima[4], Péter Udvarhelyi[5,6,7], Gergő Thiering[6], Adam Gali[6,7], Jörg Wrachtrup[1], and Florian Kaiser[1,#]

* n.morioka@pi3.uni-stuttgart.de

# f.kaiser@pi3.uni-stuttgart.de

## Affiliations

1. 3rd Institute of Physics, University of Stuttgart and Institute for Quantum Science and Technology IQST, 70569, Stuttgart, Germany
2. Advanced Research and Innovation Center, DENSO CORPORATION, 470-0111, Nisshin, Japan
3. Department of Physics, Chemistry and Biology, Linköping University, SE-58183, Linköping, Sweden
4. National Institutes for Quantum and Radiological Science and Technology, 370-1292, Takasaki, Japan
5. Department of Biological Physics, Eötvös University, Pázmány Péter sétány 1/A, H-1117 Budapest, Hungary
6. Wigner Research Centre for Physics, P.O. Box 49, H-1525 Budapest, Hungary
7. Department of Atomic Physics, Budapest University of Technology and Economics, Budafoki út 8., H-1111 Budapest, Hungary

## Supplementary Note 1: Saturation behaviour of single h-V$_{Si}$ centre under pulsed laser excitation

To evaluate the optical excitation efficiency in the pulsed regime, we use a 780 nm picosecond laser diode (PicoQuant LDH-P-C-780). The laser operation regime was kept constant to maintain the same pulse shape and duration throughout all measurements. The pulse energy $E$ was subsequently varied and we observed the resulting photon count rate $I_{\text{pulsed}}$ in the zero-phonon line, as shown in Supplementary Fig. 1.

In this experiment, we used a slow repetition rate of 2 MHz to assure that the electron comes back to the ground state before laser excitation (and is not trapped in the metastable state with ≈ 100 ns lifetime). Under this condition, we can interpret the background subtracted relative intensity as excitation probability. Since the laser pulse length (56 ps FWHM) is much shorter than the excited state lifetime of h-V$_{Si}$ centre (6 ns), the intensity saturation can be modelled by an exponential equation

$$I_{\text{pulsed}}(E) = I_{0,\text{pulsed}} \left[1 - \exp\left(-\frac{E}{E_0}\right)\right], \tag{1}$$

where $I_{0,\text{pulsed}}$ is the saturation intensity and $E_0$ is the pulse energy at which the excitation probability is $1 - e^{-1} = 0.632$. We obtained $E_0 = 4.0 \pm 0.1$ pJ from the fitting. The typical laser pulse energy for HOM experiments is 5.5 pJ, which corresponds to an excitation probability of about 74 %.

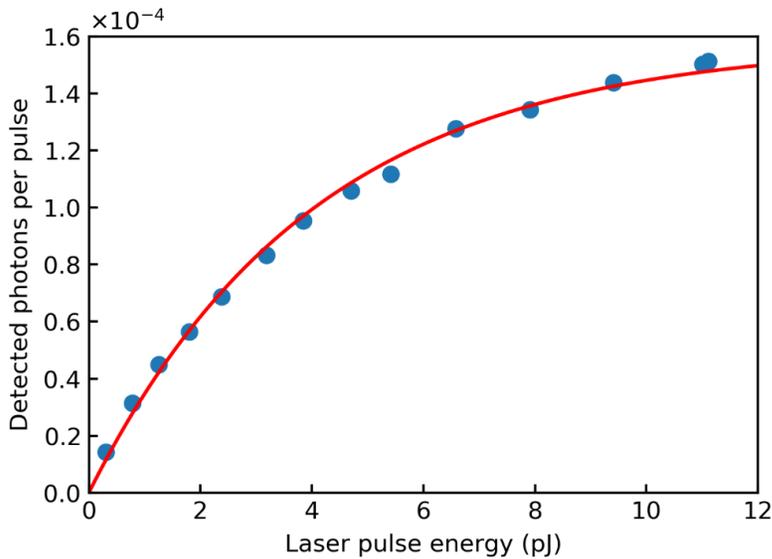

Supplementary Figure 1. Saturation behaviour of photoluminescence intensity in the zero-phonon line pulsed laser excitation at 780 nm. The pulse length is 56 ps in FWHM. The dots are experimental results and the red lines are the fitting to the model Supplementary Eq. (1), yielding a saturation pulse energy of $E_0 = (4.0 \pm 0.1)$ pJ.

## Supplementary Note 2: Evaluation of the interferometer — transmissivity/reflectivity ratio of two beam splitters and fringe contrast

In this study, an unbalanced Mach-Zehnder type fibre-based interferometer (Supplementary Fig. 2) is used for HOM type two-photon interference. To characterise the quality of the interferometer, the intensity transmissivity/reflectivity ratio ($T/R$) of the two beam splitters ($BS_1$ and $BS_2$) has to be measured. To this end, we excite a single h-$V_{Si}$ centre by the pulsed laser at 4.1 MHz repetition rate. We detect the rate of ZPL photons through the interferometer by two SNSPDs ($D_1$ and $D_2$) in early ($0 \leq t \leq 48$ ns) and late ($\delta t \leq t \leq \delta t + 48$ ns) time bins. Here, $\delta t = 48.7$ ns is the path travel time difference of two arms. As $\delta t$ is much longer than the photon coherence time, there is no single-photon interference at the output. The photons in the early time bin take the shorter arm and the photons in the late time bin take the longer arm. The integrated photon counts of two detectors in early and late time bins are $N(D_1, \text{early}) \equiv N_{11} = \eta_1 T_1 R_2 N_0$, $N(D_1, \text{late}) \equiv N_{12} = \eta_1 R_1 T_2 N_0$, $N(D_2, \text{early}) \equiv N_{21} = \eta_2 T_1 T_2 N_0$, and $N(D_2, \text{late}) \equiv N_{22} = \eta_2 R_1 R_2 N_0$, where $\eta_i$ is the detection efficiency of $D_i$ and $N_0$ is the total number of input photons. From these relationships, the $T/R$ ratios of two beam splitters are calculated to be $T_1/R_1 = \sqrt{N_{11}N_{21}/N_{12}N_{22}} = 1.129 \pm 0.006$ and $T_2/R_2 = \sqrt{N_{12}N_{21}/N_{11}N_{22}} = 1.046 \pm 0.005$.

The fringe contrast of the interferometer is measured with a highly coherent monochromatic laser (Toptica DLC DL pro 850) at ZPL wavelength of h-$V_{Si}$ centre. By optimising the polarisation rotation in the long interferometer arm with a fibre polarisation controller, the maximum interference fringe contrast obtained in this interferometer is $(1 - \varepsilon) = 0.995$. Since the theoretical limit of the fringe contrast with unbalanced $T/R$ ratios is $2\left(\sqrt{T_1T_2/R_1R_2} + \sqrt{R_1R_2/T_1T_2}\right)^{-1} = 0.9965 \pm 0.0006$, we consider the interferometer to be well aligned.

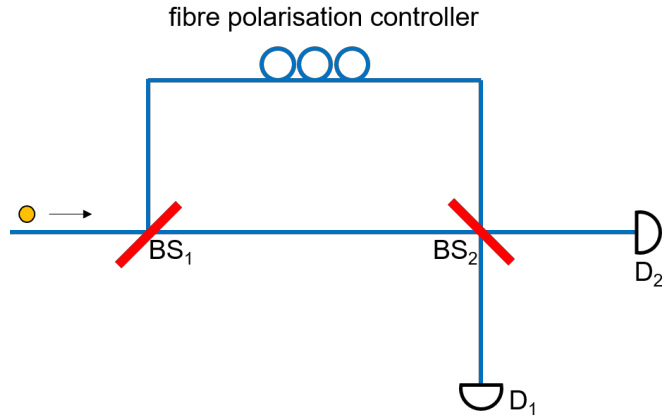

Supplementary Figure 2. Schematic image of the interferometer. $BS_{1,2}$: 50:50 beam splitter, $D_{1,2}$: superconducting nanowire single photon detectors (SNSPD).

## Supplementary Note 3: Correction of experimental imperfection factors in Hong-Ou-Mandel interference visibility

The HOM visibility gives the overlap integral of the wave packet of two photons in ideal conditions, but the experiment is affected by timing jitter of photon arrival time, background noise photons, and interferometer imperfections such as non-unity fringe contrast and unbalanced transmissivity and reflectivity. The timing jitter decreases the two-photon overlap integral. The existence of background noise photon decreases the probability of events to have two indistinguishable photons from the h-$V_{Si}$ centre at the beam splitter, resulting in the decrease of a raw HOM visibility. In this note, we extend the discussion on the HOM visibility by Santori et al.[1] by considering the effect of noise photons to estimate the correct photon overlap integral.

The signal photons are from $A_1$ and $A_2$ transitions of the h-$V_{Si}$ at the focus and the noise photons come from the ensemble of silicon vacancies on surface, bulk fluorescence, Raman scattering, laser breakthrough, etc.. We denote the probability to have a photon from the signal and noise sources per one laser pulse to be $p$ and $q$, respectively. Since the pulse length of the excitation laser (56 ps) is much shorter than the excited state lifetime of h-$V_{Si}$ (6 ns), the probability to have two ZPL photons from the same h-$V_{Si}$ centre by one excitation laser is negligibly small. The noise photons can be modelled as a Poissonian photon source, however we can safely neglect the probability to have two noise photons per laser pulse since the average number of noise photon is much smaller than 1 per laser pulse. Under these assumptions, we write the probability to have $n$ photons per laser pulse, $p_n$ ($n = 0, 1, 2$), and the signal to noise ratio SN as

$$\begin{cases} p_0 = (1-p)(1-q), \\ p_1 = p(1-q) + (1-p)q, \\ p_2 = pq, \end{cases} \quad (2)$$

$$\text{SN} = \frac{p}{q}. \quad (3)$$

Using these parameters and the other mentioned non-ideal parameters, the coincidence counts of five peaks in two-pulse HOM excitation scheme are calculated to be

$$A_0 = (p_1 + 2p_2)^2 \eta_1 \eta_2 N_0 \left\{ T_1 R_1 \left[ (T_2^2 + R_2^2) - 2\left(\frac{\text{SN}}{\text{SN}+1}\right)^2 (1-\varepsilon)^2 T_2 R_2 V \right] + 2g(T_1^2 + R_1^2) T_2 R_2 \right\}, \quad (4)$$

$$\begin{cases} A_{+1\cdot\Delta t} = (p_1 + 2p_2)^2 \eta_1 \eta_2 N_0 [(T_1^2 + R_1^2) T_2 R_2 + 2g T_1 R_1 T_2^2], \\ A_{-1\cdot\Delta t} = (p_1 + 2p_2)^2 \eta_1 \eta_2 N_0 [(T_1^2 + R_1^2) T_2 R_2 + 2g T_1 R_1 R_2^2], \end{cases} \quad (5)$$

$$\begin{cases} A_{+2\cdot\Delta t} = (p_1 + 2p_2)^2 \eta_1 \eta_2 N_0 \cdot T_1 R_1 T_2^2, \\ A_{-2\cdot\Delta t} = (p_1 + 2p_2)^2 \eta_1 \eta_2 N_0 \cdot T_1 R_1 R_2^2, \end{cases} \quad (6)$$

where $N_0$ is the number of repetitions of the experiment, $(1 - \varepsilon)$ is the interferometer's fringe contrast, and $V$ is the overlap integral of two photons from the h-$V_{Si}$ centre. The parameter $g = 2p_2/(p_1 + 2p_2)^2$ comes from the events in which two photons enter in the interferometer during one laser excitation pulse. Thus, it is equal to $g^{(2)}(\tau = 0)$ when the autocorrelation measurement is performed under the same condition as the HOM experiment. When $g^{(2)}(\tau = 0)$ degrades solely due to reduced signal-to-noise (SN), this parameter can be written as

$$g = \frac{2pq}{(p+q)^2} = \frac{2\text{SN}}{(\text{SN}+1)^2}. \tag{7}$$

This equation gives the lower limit of the parameter $g$ and $g^{(2)}(\tau = 0)$. Supplementary Fig. 3 shows the comparison of experimentally measured $g^{(2)}(\tau = 0)$ (the same data as Fig. 1(d) in main text), and the SN limited value calculated from Supplementary Eq. (7). At pulse energies below 6 pJ, $g^{(2)}(\tau = 0)$ is close to the SN limit. The degradation of $g^{(2)}(\tau = 0)$ at higher laser pulse energies is probably due to double excitation within one laser pulse. Note that in the HOM experiment, the effect of double excitation is greatly minimised by time-gating in the first several ns.

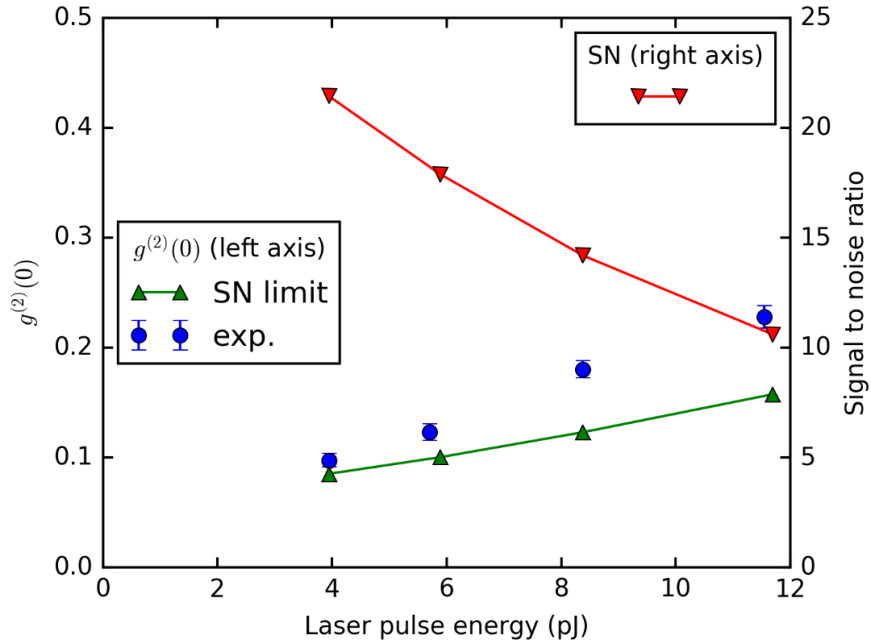

Supplementary Figure 3. $g^{(2)}(0)$ measured with pulsed excitation at repetition rate of 20.5 MHz (blue circles, left axis) and the lower bound of $g^{(2)}(0)$ (green triangles, left axis) calculated using Supplementary Eq. (7) from separately measured signal-to-noise ratio (red squares, right axis). Lines are guides to the eye. The difference between the experimentally measured $g^{(2)}(0)$ and SN limited one corresponds to $g^{(2)}(0)$ with background correction, i.e., the nonideality of the emitter.

From the experimentally obtained raw HOM visibility $V_0 = 1 - 2A_0/(A_{+1 \cdot \delta t} + A_{-1 \cdot \delta t})$, the corrected HOM visibility (two-photon overlap integral) is extracted to be

$$V = \frac{1}{(1-\varepsilon)^2}\left[\left\{\left(\frac{SN+1}{SN}\right)^2 \alpha_2 + \frac{2\alpha_1}{SN}\right\} - (1-V_0)\left\{\left(\frac{SN+1}{SN}\right)^2 \alpha_1 + \frac{2\alpha_2}{SN}\right\}\right], \tag{8}$$

$$\alpha_i = \frac{1}{2}\left(\frac{T_i}{R_i} + \frac{R_i}{T_i}\right) \quad (i = 1, 2). \tag{9}$$

Supplementary Fig. 4 shows the reduction of maximally observable experimental visibility due to finite SN ratio. Here, we assume an ideal interferometer ($\varepsilon = 0, \alpha_i = 1$) and perfect two-photon overlap integral ($V = 1$). When SN is 28, which is a typical value for this study at the laser pulse energy of 5.5 pJ, the maximum achievable HOM visibility is upper bound at 87%. Note that for HOM experiments the repetition rate is lower compared to $g^{(2)}(0)$ measurements shown in Supplementary Fig. 3.

Note that SN is a function of the time gating strategy since the noise is composed of photons from various sources with different time scales. For example, the breakthrough of the excitation laser and the Raman scattering are very fast components, but the fluorescence from ensemble of silicon vacancies on surface has the same time constant as the signal photons.

In addition to the visibility reduction due to finite SN, we now consider the effect of timing jitter of the photon arrival time. When the arrival timing of two photons at the second beam splitter is different by $\delta t_{\text{arrival}}$, the visibility decreases by a factor of $\exp(-|\delta t_{\text{arrival}}|/\tau_{\text{ES}})$.[2] By convoluting with a Gaussian distribution of timing jitter (standard deviation = $\sigma_{\text{jitter}}$), the averaged photon overlap integral decreases by a factor of

$$\beta_{\text{jitter}} = \exp\left[\left(\frac{\sigma_{\text{jitter}}}{\sqrt{2}\tau_{\text{ES}}}\right)^2\right] \text{erfc}\left(\frac{\sigma_{\text{jitter}}}{\sqrt{2}\tau_{\text{ES}}}\right). \tag{10}$$

We estimate the timing jitter of the excitation laser to be about 55 ps (one standard deviation), most of which comes from the trigger pulse generation electronics. By adding the jitter coming from the finite pulse length (56 ps in FWHM), the laser related timing jitter is estimated to be 60 ps (one standard deviation). We assume that the timing jitter caused through the ultra-fast relaxation process in the excited state vibronic levels is negligibly small, thus we take $\sigma_{\text{jitter}} = 60$ ps. As a consequence, the photon overlap integral decreases by 0.8 %. As a result, the HOM visibility after correction (including imperfection of the interferometer, SN ratio, and timing jitter) is

$$V = \frac{1}{2(1-\varepsilon)^2 \beta_{\text{jitter}}}\left[\left\{\left(\frac{SN+1}{SN}\right)^2 \alpha_2 + \frac{2\alpha_1}{SN}\right\} - (1-V_0)\left\{\left(\frac{SN+1}{SN}\right)^2 \alpha_1 + \frac{2\alpha_2}{SN}\right\}\right]. \tag{11}$$

Comparing our theoretical model with the time gated raw data shown in Fig. 4(b) in the main text, we find that the maximum achievable HOM visibility is upper bound at $(86 \pm 1)\%$ by SN, interferometer imperfections and timing jitter. The experimentally extracted visibility parameter $V = 0.85 \pm 0.04$ underlines that essentially ideal contrast could be reached by further improving the setup and noise filtering strategy.

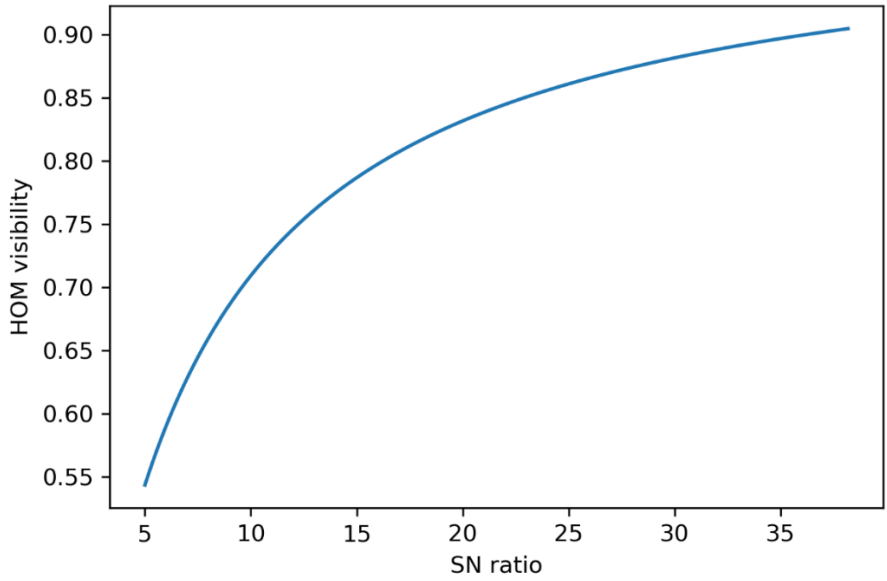

Supplementary Figure 4. Theoretical limit of raw HOM interference visibility in the two-pulse excitation scheme by taking into account the background noise (SN: signal to noise ratio) calculated with Supplementary Eq. (8). The HOM visibility is defined by $1 - 2A_0/(A_{+1\cdot\delta t} + A_{-1\cdot\delta t})$ where $A_0$ is the coincidence at zero time delay and $A_{\pm 1 \cdot \delta t}$ are those at the time delay of $\pm \delta t$. In this plot, unity two-photon overlap integral and ideal interferometer are assumed.

## Supplementary Note 4: Resonant Rabi oscillation experiment with short radiofrequency pulse for spin-controlled indistinguishable photon generation

To control the colour of photons via coherent control of the ground state spin, a radiofrequency (RF) pulse is applied to the centre between two laser excitation pulses. The first optical transition initialises the spin state into one of the Kramers doublet subspaces $m_s = \pm 1/2$ or $\pm 3/2$ depending on the observed colour of the emitted zero phonon line photon ($A_1$ or $A_2$, respectively). The RF pulse coherently manipulates the spin state, and the resulting spin population in each subspace directly translated to the probability to observe the second photon in $A_1$ or $A_2$. Considering the time difference of interferometer arms $\delta t = 48.7$ ns and the system's excited state lifetime $\tau_{ES} = 6$ ns, the allowed maximum pulse length is about 30 ns. Due to the short pulse length, we expect a frequency broadening exceeding the ground state zero-field splitting (ZFS) of 4.5 MHz. Thus, the RF field will drive all spin transitions simultaneously, leading to spin manipulation with non-unity fidelity. To determine optimal RF pulse length and the associated spin populations, we measured Rabi oscillation with resonant laser excitation.

The h-$V_{Si}$ centre is irradiated by a laser resonant to the $A_2$ ($A_1$) transition for 9 µs, which initialises the spin state into $m_s = \pm 1/2$ ($\pm 3/2$) subspace. Thereafter, a RF pulse resonant to the $|3/2\rangle_{GS} \leftrightarrow |1/2\rangle_{GS}$ transition is applied. Subsequently, the population of the spin sublevels $m_s = \pm 3/2$ ($\pm 1/2$) is read out by the same laser via the fluorescence intensity in the first 500 ns. Supplementary Figs. 5 (a) and (b) show the normalised results of Rabi experiment at RF power of 30 dBm. We find that the oscillation pattern is different from a simple cosine function, but we still observe oscillations due to spin population flips (a detailed theoretical model of this Rabi result will be given later in this Supplementary Note). We determine three relevant pulse durations: 10 ns to induce a $\pi/4$-pulse, 19 ns to induce a $\pi/2$-pulse, and 29 ns to induce a $3\pi/4$-pulse.

Due to the high-power RF condition the sample heats up significantly, which causes optical line broadening. Supplementary Fig. 5(c) shows typical optical absorption spectra (photoluminescence excitation spectroscopy) measured under similar RF conditions as used during the HOM experiments. To this end, we apply RF pulses with 30 dBm power at a repetition cycle of $10 \cdot \delta t = 487$ ns. For RF pulse durations of 10 ns, 19 ns, and 29 ns, we observe linewidths of $82 \pm 1$ MHz, $86 \pm 1$ MHz, $112 \pm 2$ MHz. Although not implemented here, we note that this heating can be minimised by optimising the structure of the RF antenna to obtain a better RF field coupling to the spin.

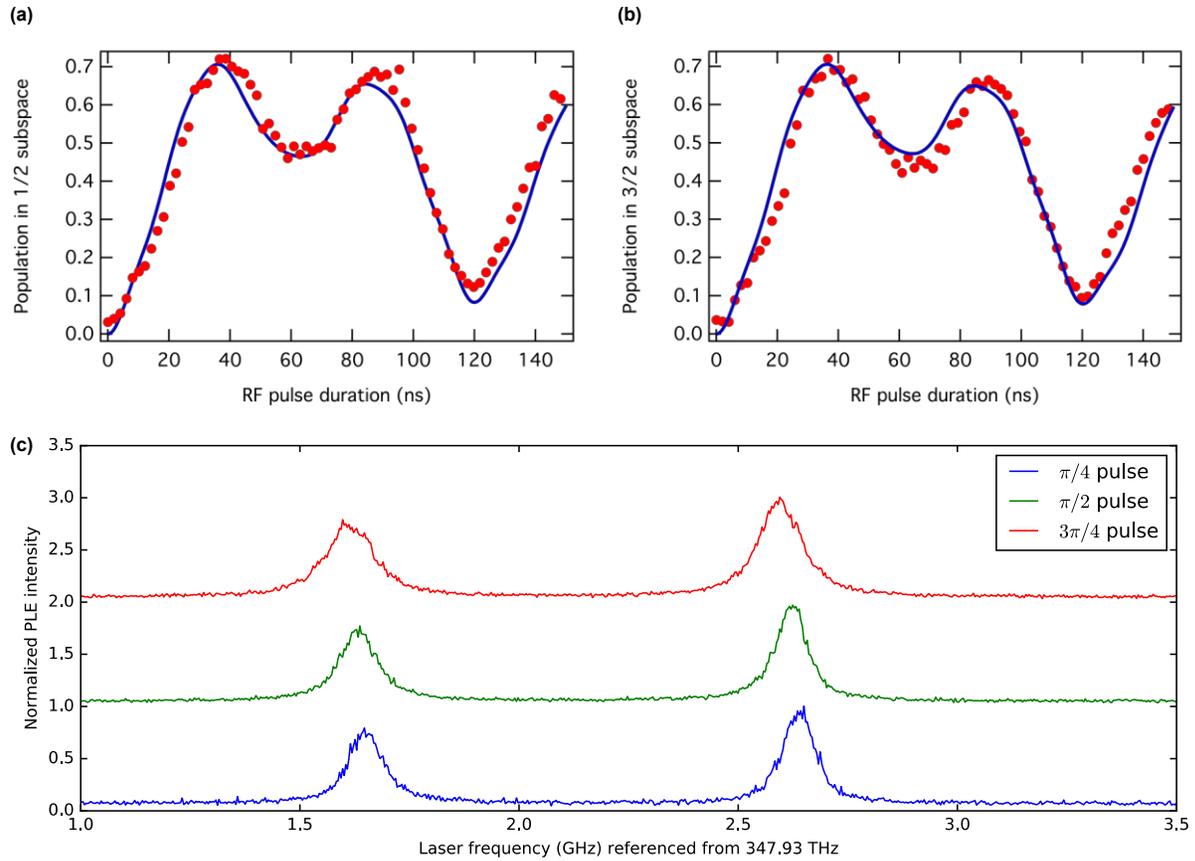

Supplementary Figure 5. Typical experimental data for Rabi oscillations under resonant excitation and strong RF driving field. (a) Spin population development after the system is initialised into the $m_s = \pm 1/2$ subspace. (b) Spin population development after the system is initialised into the $m_s = \pm 3/2$ subspace. Red dots are data, and blue lines are fits. (c) Optical absorption spectra measured by photoluminescence excitation spectroscopy under the existence of RF pulses at the repetition cycle of 487 ns. The laser power was set to 3.5 W/cm² to reduce laser power broadening. For the $\pi/4, \pi/2$ and $3\pi/4$ pulses, the observed linewidths are $82 \pm 1$ MHz, $86 \pm 1$ MHz, $112 \pm 2$ MHz, respectively (average over both transitions).

For proper interpretation of the visibility of the spin-controlled HOM interference experiments, we quantify the maximally achievable HOM visibility at those different RF pulse and temperature conditions. To this end, we always perform an additional HOM interference experiment in which the identical RF pulse is applied right before an experimental sequence, instead of applying it during the sequence. A typical measurement with uncorrected data is shown in Supplementary Fig. 6. Then we obtain the normalised HOM visibility as

$$V_{\text{norm}} = \frac{V(\text{RF during sequence})}{V(\text{RF before squence})}. \quad (12)$$

Here, $V(\text{RF during sequence})$ and $V(\text{RF before squence})$ are corrected for experimental imperfections by Supplementary Eq. (11). This way, the value $(1 - V_{\text{norm}})$ equals to the amount of spin population flip from the initial spin subspace to the other, which can be directly compared with Rabi oscillation result as shown in Fig. 3(d) in the main text.

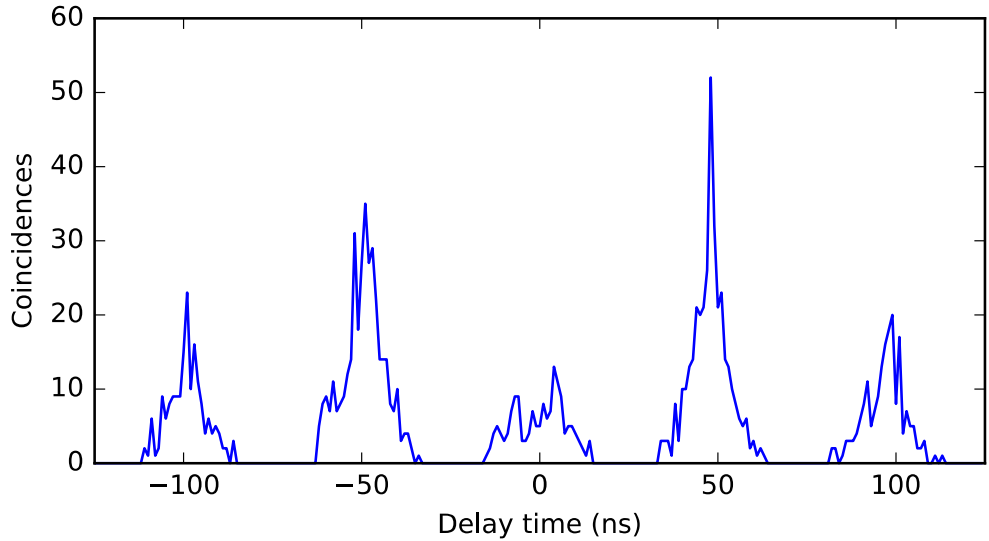

Supplementary Figure 6. Reference HOM measurement with $\frac{\pi}{2}$ pulse before the experimental sequence. Time gating settings are $t_{\text{Start}} = 2$ ns and $\Delta t = 16.5$ ns. The observed raw HOM visibility is $0.56 \pm 0.04$. After experimental imperfection corrections, we obtain $V(\text{RF before sequence}) = 0.65 \pm 0.05$.

## Time evolution of ground state spin populations under pulsed strong RF drive

To evaluate the time-dependent spin populations in the ground states under strong RF drive, we start with the static Hamiltonian, describing the system in an external magnetic field aligned along the z-axis:

$$H_0 = D \cdot S_z^2 + \gamma_e \cdot B_z \cdot S_z. \quad (13)$$

Here, $D = 2\pi \cdot 2.25$ MHz is the ground state zero-field splitting, $\gamma_e = 2\pi \cdot 28$ GHz/T the electron gyromagnetic ratio, and $B_z \approx 0.9$ mT the externally applied field. Our RF drive is modelled by the interaction Hamiltonian:

$$H_{\text{RF}} = \Omega \cdot \cos(2\pi f t + \phi) \cdot S_x \quad (14)$$

Here, $\Omega \approx 2\pi \cdot 14$ MHz is the strength of the RF driving field, $f = 30.26911$ MHz is the RF frequency, $t$ the time, and $\phi$ is a (random) phase term, which accounts for the fact that the RF driving is here faster than the ground state level separation $(2 \cdot D)$ and not phase-synchronised, thus the rotating wave approximation might not be valid. With these Hamiltonian operators, the time evolution operator is given by

$$U(t) = e^{-i \int_0^t [H_0 + H_{\text{RF}}(t')] dt'}. \quad (15)$$

The time evolution of the four spin states in the ground state is then

$$\rho(t) = U(t) \rho_0 U^\dagger(t). \quad (16)$$

Here, $\rho(t)$ is the density matrix describing the system at time $t$ and $\rho_0$ is the state at $t = 0$. We obtain $\rho(t)$ by solving Supplementary Eq. (15) numerically, averaging over the random phase term $\phi$, and using small time steps $dt \approx 0.5$ ns, which is significantly smaller than the typical time scale of spin state development ($\approx 10$ ns).

For the spin-controlled HOM interference experiments, emission of the first photon in the $A_1$ or $A_2$ line projects the state into the $m_s = \pm\frac{1}{2}$ or $m_s = \pm\frac{3}{2}$ ground state spin subspace, respectively. Thus, we have to distinguish between two realisations in which $\rho_0 = \frac{1}{2} \left( \left|\frac{1}{2}\right\rangle \left\langle\frac{1}{2}\right| + \left|-\frac{1}{2}\right\rangle \left\langle-\frac{1}{2}\right| \right) \equiv \rho_0^{(1/2)}$ or $\rho_0 = \frac{1}{2} \left( \left|\frac{3}{2}\right\rangle \left\langle\frac{3}{2}\right| + \left|-\frac{3}{2}\right\rangle \left\langle-\frac{3}{2}\right| \right) \equiv \rho_0^{(3/2)}$, respectively. Supplementary Figs. 5 (a) and (b) shows the comparison of our theoretical model with the experimental results of resonant Rabi oscillation (described in Supplementary Note 4) with the initial state of $\rho_0^{(3/2)}$ and $\rho_0^{(1/2)}$, respectively. To fit the data to our model, we normalise the amplitude of both the experimental data and theoretical model. We solve Supplementary Eq. (16) with two free parameters, i.e. $\Omega$ and $B_z$, and minimise error squares. From both data sets, we extract a Rabi frequency of $\Omega = 2\pi \cdot (14.4 \pm 0.2)$ MHz and $B_z = 0.919 \pm 0.003$ mT. In the end, we correct both the experimental data and the theoretical model for the above-used amplitude normalisation factor.

In the HOM experiment under off-resonant excitation, each individual initial spin-state subspace is randomly chosen with almost equal probability after the system experiences intersystem crossing[3]. The resulting HOM visibility is an average of the cases with two different initial states. However, the data and fits show that the spin population transfer from each spin subspace to the other is almost equally efficient. Therefore, we can directly model the normalised HOM visibility as $V_{\text{norm}} = 1 - \bar{p}$, in which $\bar{p}$ is the amount of flipped spin population, after averaging over both realisations (i.e. with the system being in the initial state of $\rho_0^{(3/2)}$ and $\rho_0^{(1/2)}$, respectively). The associated model is plotted as a solid line in Fig. 3(d) in the main text.

## Supplementary Note 5: Analysis of pure dephasing rate and spectral diffusion amplitude by HOM visibility with time-gating

First, we derive the Eq. (1) in the main text. We follow the derivation given by Thoma et al.[4]. Assume the photons arrive at the beam splitter at $t = 0$ and detected at $t = t_D$ and $t = t_D + \tau$ at different detectors. The time-resolved coincidence count rate per a pair of photons is given by

$$G^{(2)}(t_D, \tau) = \Gamma^2 (1 - e^{-\gamma \tau}) e^{-\Gamma(2t_D + \tau)}. \tag{17}$$

Here, $\Gamma = \frac{1}{6 \text{ ns}}$ is the inverse excited state lifetime, and $\gamma = \Gamma_0'[1 - e^{-(\delta t/\tau_c)^2}] + 2\gamma'$, with $\Gamma_0'$ being the amplitude of spectral diffusion, $\tau_c$ the associated time constant, and $\gamma'$ the pure dephasing rate of the single emitter[5]. Also, we consider ideal 50:50 beam splitter since we compare this theory with experimental HOM visibility after imperfection correction by Supplementary Eq. (11). Integration of Supplementary Eq. (17) over $t_D$ and $\tau$ within the gated detection time window $[t_{\text{Start}}, t_{\text{Stop}}]$ gives the HOM visibility after normalisation. The coincidence for normalisation is given by

$$G'^{(2)}(t_D, \tau) = \Gamma^2 e^{-\Gamma(2t_D + \tau)}. \tag{18}$$

Therefore, the HOM visibility with time gating analysis is calculated to be

$$V = 1 - \frac{\int_{t_{\text{Start}}}^{t_{\text{Stop}}} dt_D \int_0^{t_{\text{Stop}} - t_D} d\tau G^{(2)}(t_D, \tau)}{\int_{t_{\text{Start}}}^{t_{\text{Stop}}} dt_D \int_0^{t_{\text{Stop}} - t_D} d\tau G'^{(2)}(t_D, \tau)}, \tag{19}$$

which equals to Eq. (1) in the main text.

By fitting the data in main text Fig. 4(c) with the model in Eq. (1), we extract $\gamma$ (which is the only free fitting parameter). For further analysis, we decompose $\gamma$ into a slow term (related to spectral diffusion with amplitude $\Gamma_0'$), and a fast term (related to pure dephasing with rate $\gamma'$). As outlined in the main text, the major contribution to HOM visibility reduction is pure dephasing, while spectral diffusion due to laser ionisation is relatively slow compared to the experimental time scale, i.e. $\tau_c \gg \delta t = 48$ ns. Then we can directly extract the maximum pure dephasing rate, as it is $\gamma = 2\gamma'_{\text{max}}$. In other words, HOM visibility reduction is only due to pure dephasing. The amplitude of (slow) spectral diffusion $\Gamma_0'$ is obtained by considering the emission linewidth (measured over time scales of seconds to minutes). The FWHM optical linewidth is

$$\Delta \nu = \frac{\Gamma + \Gamma_0' + \gamma'}{2\pi}. \tag{20}$$

For reasons of completeness, we can also assume the opposite scenario, *i.e.* no pure dephasing ($\gamma' = 0$ MHz), such that HOM visibility contrast reduction is solely explained by fast spectral diffusion. In this case we compute the maximum spectral diffusion amplitude $\Gamma'_{0,\text{max}}$ according to Supplementary Eq. (20), and solve $\gamma = \Gamma'_0 \left[1 - e^{-(\delta t/\tau_{c,\text{min}})^2}\right]$ for the minimal spectral diffusion time constant $\tau_{c,\text{min}}$.

The related results are shown in the in Supplementary Table I.

Supplementary Table I. Summary of temperature dependent spectral linewidth (averaged over both transitions $A_1$ and $A_2$), the maximum pure dephasing rate $\gamma'_{\text{max}}$ and associated spectral diffusion amplitude $\Gamma'_0$ for the model in which the HOM contrast is limited by pure dephasing. Additionally, we give the maximum amplitude of spectral diffusion $\Gamma'_{0,\text{max}}$ and its minimum time constant $\tau_{c,\text{min}}$ for the model in which the HOM contrast is solely limited by spectral diffusion (*i.e.* $\gamma' = 0$).

| Temperature [K] | PLE linewidth [MHz] | Pure dephasing limited | | Spectral diffusion limited | |
|---|---|---|---|---|---|
| | | $\gamma'_{\text{max}}/2\pi$ [MHz] | $\Gamma'_0/2\pi$ [MHz] | $\Gamma'_{0,\text{max}}/2\pi$ [MHz] | $\tau_{c,\text{min}}$ [ns] |
| 5.0 | $62.4 \pm 0.4$ | $3.2 \pm 0.4$ | $32.7 \pm 0.6$ | $35.9 \pm 0.4$ | $109 \pm 8$ |
| 5.9 | $70.1 \pm 0.3$ | $6.7 \pm 0.8$ | $36.9 \pm 0.9$ | $43.6 \pm 0.3$ | $81 \pm 6$ |
| 6.8 | $82.4 \pm 0.3$ | $16.6 \pm 2.4$ | $39.3 \pm 2.4$ | $55.9 \pm 0.3$ | $51 \pm 6$ |

## Supplementary Note 6: Vibronic interaction theory

The origin of dephasing in the optical signal for V1 center is the coupling to the V1' polaronic excited state mediated by acoustic phonons. As outlined in more details in Udverhelyi et al.[6], at very low temperatures, only the acoustic phonons have significant occupation number. However, compared to the temperatures in the experiment, the polaronic gap between V1 and V1' is relatively large (4.4 meV), which excludes the consideration of two-phonon (Raman scattering) processes to be competitive with the single phonon absorption process. Thus we describe the dephasing with a resonant phonon coupling[7]. This can be formulated using time-dependent perturbation theory with first order contribution in the linear vibronic interaction leading to Fermi's Golden Rule formula for the transition rate

$$\gamma' = 2\pi \sum_k n_k |\chi_k|^2 \delta(\Delta E - \hbar\omega_k), \qquad (21)$$

where $k$ is the index of phonon mode, $n_k$ is the acoustic phonon occupation number, $\chi_k$ is the linear vibronic interaction strength, $\hbar\omega_k$ is the acoustic phonon energy, and $\Delta E$ is the energy gap between V1 and V1' levels. For the density of acoustic phonon states $\rho(\omega)$ we use the Debye-model as $\rho(\omega) = \rho\omega^2$, where $\rho$ is a constant. We can approximate $\overline{|\chi_k|^2} \approx \chi\omega$ phonon mode average for the acoustic phonons, where $\chi$ is a constant. After this, the summation results in

$$\gamma' = \frac{2\pi}{\hbar^3} \rho\chi(\Delta E)^3 n(\Delta E, T), \qquad (22)$$

where we use the thermal occupation function of phonons $n(\Delta E, T)$. Since $\Delta E$ is relatively large we find the low temperature limit of this function with exponential temperature dependence, as described in Eq. (2) in the main text.

## Supplementary Note 7: Analysis of quantum beating with spin control via RF pulse

This note explains the analysis of quantum beating obtained with the HOM interference experiment with spin-flip RF pulses. The time delay of the RF pulse from the first laser pulse is 18 ns. Therefore, the coincidence data is taken within the detection time window $[t_{\text{Start}}, t_{\text{Stop}}] = [1.5 \text{ ns}, 18 \text{ ns}]$ with the time-gating technique described in Fig. 4(a) in the main text. This strategy rejects the laser related noise and ensures that the system is in the ground state while RF pulse is applied for the collected data. Due to non-unity spin flip fidelity in our conditions and the existence of noise photons, we consider three components in the coincidence data. The first component is the interference of photons from different transitions $\{A_1, A_2\}$, which causes beating due to the frequency difference of two photons $\delta v \cong 1$ GHz. The coincidence count distribution per pair of two photons (detected at $t = t_D$ and $t = t_D + \tau$ at different detectors, $t = 0$: the earliest possible arrival time) for this case is given by[4,5]

$$G_1^{(2)}(t_D, \tau) = \Gamma^2 [1 + \cos(2\pi\delta v \tau + \pi) e^{-\gamma\tau}] e^{-\Gamma(2t_D + \tau)}, \tag{23}$$

where $\Gamma = \frac{1}{6 \text{ ns}}$ is the inverse excited state lifetime, $\gamma$ is the sum of the pure dephasing rate and the spectral diffusion rate discussed in Supplementary Note 5. The integration of the Supplementary Eq. (23) within the detection time window $t_D \in [t_{\text{Start}}, t_{\text{Stop}} - \tau]$ gives the beating pattern in the correlation measurement

$$\bar{G}_1^{(2)}(0 < \tau < \Delta t) = \int_{t_{\text{Start}}}^{t_{\text{Stop}} - \tau} dt_D \, G_1^{(2)}(t_D, \tau)$$

$$= \frac{\Gamma}{2} e^{-2\Gamma t_{\text{Start}}} e^{-\Gamma\tau} [1 - e^{-2\Gamma(\Delta t - \tau)}][1 + \cos(2\pi\delta v \tau - \pi) e^{-\gamma\tau}]. \tag{24}$$

Here, $\Delta t = 16.5$ ns is the detection window. The second component is the interference of photons from the same transition $\{A_i, A_i\}$ ($i = 1, 2$), which results from non-perfect spin flip fidelity by RF pulse. The corresponding correlation data $\bar{G}_2^{(2)}(\tau)$ is obtained by substituting $\delta v = 0$ to Supplementary Eq. (24)

$$\bar{G}_2^{(2)}(\tau) = \frac{\Gamma}{2} e^{-2\Gamma t_{\text{Start}}} e^{-\Gamma\tau} [1 - e^{-2\Gamma(\Delta t - \tau)}][1 - e^{-\gamma\tau}]. \tag{25}$$

For simplicity, we assume $\gamma$ in Supplementary Eqs. (24) and (25) are approximated to be the same. The third component is the interference involves noise photons and remained coincidences resulting from the imperfection of the interferometer. Here, we assume that noise photons have the same decay time as the signal photon, which is indeed observed in the experiment and is probably from an h-$V_{\text{Si}}$ ensemble that exists on the surface. The fast noise photons (laser related) are filtered out by time-gating. The frequency difference of photons involving noise photons is randomly distributed and the optical coherence time is expected to be much shorter than the timing resolution of our detection system and electronics (0.4 ns). Therefore, the corresponding coincidence counts $\bar{G}_3^{(2)}$ is obtained from Supplementary Eq. (24) by averaging cosine term and taking a limit of $\gamma \to \infty$ as

$$\bar{G}_3^{(2)}(\tau) = \frac{\Gamma}{2} e^{-2\Gamma t_{\text{Start}}} e^{-\Gamma \tau} \left[1 - e^{-2\Gamma(\Delta t - \tau)}\right]. \quad (26)$$

The total coincidence counts $\bar{G}_{\text{tot}}^{(2)}$ are obtained by summing up these three components $\bar{G}_i^{(2)}$ with associated coefficients $c_i$. By considering $\tau < 0$ region, a small time difference between two detectors $t_0$, and the finite detection timing resolution of the detectors and electronics (approximated by an Gaussian broadening with a standard deviation of $\sigma_{\text{det}}$), the total coincidence is calculated to be

$$\bar{G}_{\text{tot}}^{(2)}(\tau) = \left[\sum_{i=1}^{3} c_i \bar{G}_i^{(2)}(|\tau - t_0|)\right] * \left[\frac{1}{\sqrt{2\pi}\sigma_{\text{det}}} \exp\left(-\frac{\tau^2}{2\sigma_{\text{det}}^2}\right)\right]. \quad (27)$$

This function is fitted to the experimental data which is measured with 0.1 ns bin width and is smoothened by three-point averaging (the time binning and 3 point average are also considered in the fitting). In Supplementary Eq. (23), the non-ideal transmissivity/reflectivity ratio of the beam splitters and the non-unity interferometer fringe contrast is neglected. However, these imperfections are considered in the evaluation of the coefficient $c_3$.

We can extract the parameters $\gamma$, $c_2/c_1$, and $c_3/c_1$ from experiments. $\gamma = (119 \pm 5)$ MHz is obtained from the HOM visibility when the RF pulse is applied prior to the first laser excitation pulse (the experiment without spin flip). The coefficient ratio $c_2/(c_1 + c_2)$ corresponds to the amount of the spin population transfer by the RF pulse, which is equal to $(1 - V_{\text{norm}}) = 0.39 \pm 0.1$ as discussed in Supplementary Note 4. As a result, we obtain

$$\frac{c_2}{c_1} = \frac{V_{\text{norm}}}{1 - V_{\text{norm}}} = 1.55 \pm 0.47. \quad (28)$$

The coefficient ratio $c_3/(c_1 + c_2)$ can be evaluated from Supplementary Eq. (4)

$$\frac{c_3}{c_1 + c_2} = \frac{A_0|_{V=1}}{A_0|_{V=0} - A_0|_{V=1}} = \frac{\alpha_2 + 2\alpha_1 g}{[SN/(SN+1)]^2 (1-\varepsilon)^2} - 1 \quad (29)$$

as $A_0|_{V=1}$ corresponds to the coincidence count remained when the photons from the emitter is perfectly indistinguishable, *i.e.*, contributions from noise photons and the interferometer imperfections, and the rest is the possible number of events related to the interference of photons from $\{A_i, A_j\}$ transitions. In combination with Supplementary Eq. (29), we obtain $c_3/c_1 = 0.60 \pm 0.12$. With these fixed parameters, there are only four fitting parameters left: $c_1 = 739 \pm 11$, $t_0 = (0.00 \pm 0.02)$ ns, $\delta \nu = (0.966 \pm 0.007)$ GHz, and $\sigma_{\text{det}} = (0.16 \pm$

0.02) ns. The beating frequency $\delta\nu$ well agrees with the frequency difference of $A_1$ and $A_2$ transitions and the detection timing resolution $\sigma_{\text{det}}$ is reasonable considering the SNSPD manufacturer's specifications.

## Supplementary References